\documentclass[12pt,preprint]{aastex}
\usepackage{epsfig,graphicx}

\newcommand{\etal}{\emph{et al.}}

\begin{document}
%%%%%%%%%%%%%%%%%%%%%%%%%%%%%%%%%%%%%%%%%%%%%%%%%%%%%%%%%%%%%%%%%%%%%%%%%%%%%%%
\title{Diffusion of Cosmic Rays in the Expanding Universe. \\
II. Energy Spectra of Ultra-High Energy Cosmic Rays}
\author{V. Berezinsky}
\affil{INFN - Laboratori Nazionali del Gran Sasso, I--67010
Assergi (AQ), Italy; \\
Institute for Nuclear Research of RAS, 60th October Revolution
prospect 7a,\\ 117312 Moscow, Russia}

\and

\author{A. Z. Gazizov}
\affil{B.I. Stepanov Institute of Physics, \\
Independence Avenue 68, BY-220072 Minsk, Republic of Belarus; \\%
INFN - Laboratori Nazionali del Gran Sasso, I--67010 Assergi (AQ),
Italy}

%%%%%%%%%%%%%%%%%%%%%%%%%%%%%%%%%%%%%%%%%%%%%%%%%%%%%%%%%%%%%%%%%%%%%%%
\begin{abstract}
We consider the astrophysical implications of the diffusion
equation solution for Ultra-High Energy Cosmic Rays (UHECR) in the
expanding universe, obtained in paper~I \citep{BG06}. The UHECR
spectra are calculated in a model with sources located in vertices
of the cubic grid with a linear constant (source separation) d.
The calculations are performed for various magnetic field
configurations ($B_c, l_c$), where $l_c$ is the basic scale of the
turbulence and $B_c$ is the coherent magnetic field on this scale.
The main purpose of these calculations is to demonstrate the
validity of the solution obtained in paper~I and to compare this
solution with the Syrovatsky solution used in previous works. The
Syrovatsky solution must be necessarily embedded in the static
cosmological model. The formal comparison of the two solutions
with all parameters being fixed identically reveals the
appreciable discrepancies between two spectra. These discrepancies
are less if in both models the different sets of the best-fit
parameters are used.

\end{abstract}
%%%%%%%%%%%%%%%%%%%%%%%%%%%%%%%%%%%%%%%%%%%%%%%%%%%%%%%%%%%%%%%%%%%%%%%


%%%%%%%%%%%%%%%%%%%%%%%%%%%%%%%%%%%%%%%%%%%%%%%%%%%%%%%%%%%%%%%%%%%%%%%%%%%%%%%
\section{Introduction}
\label{sec:introduction}

Diffusive propagation of ultra-high energy cosmic rays (UHECR) in
extragalactic space has been recently studied by %
\cite{AB04,AB05,Lemoine,Aloisioetal} using the Syrovatsky
 solution (see \cite{Syrov}) of the diffusion equation. This solution
has been obtained under rather restrictive assumptions that
diffusion coefficient $D(E)$ and energy losses $b(E) = -dE/dt$ of
the propagating particles do not depend on time $t$. In our recent
work (\cite{BG06}, to be cited below as paper~I) we found the
analytic solution to the diffusion equation in the expanding
universe, valid for the time-dependent diffusion coefficient
$D(E,t)$ and energy losses $b(E,t)$. The aim of this work is to
calculate spectra of UHE protons using the BG solution of paper~I
and to compare them with the spectra obtained with the help of the
Syrovatsky solution in the above-cited papers.

The  diffusion equation for ultra-relativistic particles
propagating in the expanding universe from a single source, as
obtained in paper~I, reads
\begin{equation} %
\frac{\partial n}{\partial t} - b(E,t)\frac{\partial n} {\partial
E}+3H(t)n - n\frac{\partial b(E,t)}{\partial E} -
\frac{D(E,t)}{a^2(t)} \ \mathbf{\nabla}_x^2 n =
\frac{Q_s(E,t)}{a^3(t)} \ \delta^3(\vec{x}-\vec{x}_g),
\label{diff-basic} %
\end{equation} %
where the coordinate $\vec{x}$ corresponds to the comoving
distance and $a(t)$ is the scaling factor of the expanding
universe, $n = n(t,\vec{x},E)$ is the particle number density per
unit energy in an expanding volume of the universe, $dE/dt =
-b(E,t)$ describes the total energy losses, which include
adiabatic $H(t)E$ and interaction $b_{int}(E,t)$ energy losses.
$Q_s(E,t)$ is the generation function, that gives the number of
particles generated by a single source at coordinate $\vec{x}_g$
per unit energy and unit time.

According to paper~I, the spherically-symmetric solution of
Eq.~(\ref{diff-basic}) is
\begin{equation} %
n(x_g,E)=\int_0^{z_g} dz \left |\frac{dt}{dz}(z) \right | \
Q_s[E_g(E,z),z] \ \frac{\exp[-x_g^2/4\lambda(E,z)]}
{[4\pi\lambda(E,z)]^{3/2}} \ \frac{dE_g}{dE}(E,z),
\label{solution} %
\end{equation}%
where
\begin{equation} %
\frac{dt}{dz}(z) = - \frac 1 {H_0 \ (1+z)\sqrt{\Omega_m(1+z)^3 +
\Lambda}}
\label{dtdz(z)} %
\end{equation} %
with  cosmological parameters $\Omega_m = 0.27$ and $\Lambda =
0.73$,
\begin{equation} %
\lambda (E,z)= \int_0^z dz' \left |\frac{dt'}{dz'} \right | \
\frac{D(\mathcal{E}',z')}{a^2(z')} , %
\label{lambda(E,z)} %
\end{equation} %
\begin{equation} %
\frac{dE_g}{dE}(E,z) = (1+z)\exp\left [ \int_0^z dz'\left
|\frac{dt'}{dz'} \right | \ \frac{\partial
b_{int}(\mathcal{E}',z')}{\partial \mathcal{E}'} \right ].
\label{dEg/dE} %
\end{equation} %
The characteristic trajectory, $\mathcal{E}'=E'(E,z')$, is a
solution of the differential equation
\begin{equation} %
\frac{dE}{dt}= - [ H(t)E + b_{int}(E,t)].
\label{dEdtB} %
\end{equation} %
It gives the energy $E'$ of a particle at epoch $z'$, if this
energy is $E$ at $z=0$; we shall  use also the notation $E_g(E,z)$
for this quantity. The upper limit $z_g$ in the integral of
Eq.~(\ref{solution}) is provided by maximum energy of acceleration
as $E_g(E,z_g)=E_{\rm max}$, or by $z_{\rm max}$ whichever is
smaller.

To calculate the diffuse flux of UHE protons $J_p(E)$, we sum up
contributions of sources located in the vertices of a 3D cubic
lattice with spacing $d$. Positions of the sources
$\vec{x}_g=(\xi,\eta,\zeta)$ are given by the coordinates
$\xi=d(i+1/2)$, $\eta=d(j+1/2)$, $\zeta=d(k+1/2)$, where
$i,j,k=0,\pm 1,\pm 2,..$ and position of the observer is assumed
at $\xi=0, \eta=0, \zeta=0$. Thus we obtain
\begin{equation} %
J_p(E)=\frac{c}{4\pi} \sum_{i,j,k}n(x_{ijk},E),
\label{J_p} %
\end{equation} %
where $n(x_{ijk},E)$ is given by Eq.~(\ref{solution}) and
\begin{equation} %
x_{ijk}= d \sqrt{(i+1/2)^2 + (j+1/2)^2 + (k+1/2)^2}.
\label{x_ijk} %
\end{equation} %

For the propagation of UHE protons in magnetic fields we follow
the picture used by \cite{AB04,AB05}, namely we assume the
turbulent magnetized plasma. Magnetic field produced by turbulence
is characterized by the value of the coherent magnetic field $B_c$
on the basic scale of turbulence $l_c$, which we shall  keep in
our estimates as $1$~Mpc. On the smaller scales $l < l_c$ the
magnetic field is determined by the turbulence spectrum. The
critical energy of propagation is determined by the relation
$r_L(E_c)=l_c$, where $r_L$ is the Larmor radius of a proton.
Numerically, $E_c=0.93 \times 10^{18} (B_c/1$~nG)~eV. The
characteristic propagation length in magnetic field is the
diffusion length $l_d(E)$. It is defined as the distance on which
a particle is deflected by 1 rad. The diffusion coefficient is
defined as $D(E)= c l_d(E)/3$. For the case $r_L(E) \gg l_c$,
i.e.\ when $E \gg E_c$, the diffusion length can be
straightforwardly found from multiple scattering as
\begin{equation} %
l_d(E) = 1.2 \ \frac{E^2_{18}}{B_{\rm nG}} ~ {\rm Mpc},
\label{l_d} %
\end{equation} %
where $E_{18} = E/(10^{18}$~eV) and $B_{\rm nG}=B/(1$~nG). At
$E=E_c$,~ $l_d=l_c$.

At $E \ll E_c$ the diffusion length depends on the spectrum of
turbulence. For the Kol\-mo\-gorov spectrum $l_d (E) = l_c
(E/E_c)^{1/3}$; for the Bohm regime $l_d(E) = l_c \ (E/E_c)$.

The strongest observational upper limit on the magnetic fields in
our picture is given by \cite{BBO99} as $B_c \lesssim 6$~nG on the
scale $l_c = 50$~Mpc. In the calculations below we assume the
representative values of $B_c$ in the range $(0.1 - 1)$~nG for
$l_c = 1$~Mpc.

We do not put as the aim of this paper the detailed study of
diffusion in the time-dependent regime. Such a work \citep{ABG07}
is at present in progress. Here we want mainly to demonstrate the
validity of the solution for expanding universe obtained in paper
I and to perform the numerical comparison of the UHECR diffuse
spectra predicted by BG and Syrovatsky solutions. The difference
is expected to be substantial at energies $E \lesssim 3\times
10^{18}$~eV, where effects  of the universe expansion (in
particular, of the CMB temperature growth with red-shift) are not
negligible. We want also to test the new solution obtained in
paper~I, namely to see the compatibility of the BG and Syrovatsky
spectra at high energies, where the universe can be considered as
the static one, as well as the convergence of BG spectra to the
universal spectrum, when the source separation $d \rightarrow 0$.

The paper is organized as follows: Section~\ref{sec:whydiffusion}
addresses the question how reliable is an assumption of the
diffusion for the low-energy part of UHECR. In
Section~\ref{sec:expandingU} we calculate the diffuse spectra
using the time-dependent BG solution. In Section~\ref{sec:staticU}
we consider the static universe, which is the necessary assumption
for the Syrovatsky solution, and in Section~\ref{sec:comparison}
we compare the spectra calculated for the expanding and static
universes. The short conclusions are presented in
Section~\ref{sec:conclusions}.

%%%%%%%%%%%%%%%%%%%%%%%%%%%%%%%%%%%%%%%%%%%%%%%%%%%%%%%%%%%%%%%%%%%%%%%%%%%%%
\section{Why diffusion?}
\label{sec:whydiffusion} %
We argue here that at least at the low-energy end of extragalactic
UHECR the diffusion propagation is unavoidable for any reasonable
magnetic field. We estimate also the magnetic field $B_c$ for
which protons with energies $E \geq 10^{17}$~eV propagate
quasi-rectilinearly, hence producing the same energy spectrum as
in the case of rectilinear propagation.

To facilitate the calculations, let us consider the case of the
static universe as in \cite{AB05}, namely the universe with 'age'
$t_0=H_0^{-1}$, as follows from WMAP observations, and with
fictitious 'adiabatic' energy loss of particles $b(E) = E\, H_0$,
where $H_0=72$~km/sMpc is the observed Hubble parameter.

In this picture there is a maximum diffusive distance -- {\em
magnetic horizon} -- which is determined by the distance traversed
by a particle during the age of the universe $t_0$:
\begin{equation} %
r_{\rm hor}^2(E)= \int_0^{t_0}dt \ D[E_g(E,t)]\ ,
\label{r_hor1} %
\end{equation} %
where $E_g(E,t)$ is the energy that a particle has at time $t$, if
it is $E$ at $t=t_0$. Putting $dt = -dE_g/b(E_g)$  in
Eq.~(\ref{r_hor1}), we obtain
\begin{equation} %
r_{\rm hor}^2(E) =\int_E^{E_{\rm max}}\frac{dE_g}{b(E_g)}\ D(E_g)
\ ,
\label{r_hor} %
 \end{equation} %
where $E_{\rm max}= {\rm min}[E_g(E,t_0), E^{\rm acc}_{\rm max}]$.

In $r_{\rm hor}^2(E)$ one can recognize (see \cite{AB05}) the
Syrovatsky variable $\lambda (E,E_g)$ at $E_g=E_{\rm max}$ (for
the physical discussion of magnetic horizon see \cite{Parizot04}).

Let us consider a transition from the diffusive to the rectilinear
propagation, allowing the considerable deflection angle $\theta
 \gtrsim 1$, when the spectrum is the same as in the rectilinear
propagation. Two conditions must be fulfilled:
\begin{equation} %
r_L(E) > l_c,
\label{larm} %
\end{equation} %
\begin{equation} %
l_d(E) > r_{\rm hor}(E). %
\label{horizon} %
\end{equation} %

Eq.~(\ref{larm}) gives a necessary (but not a sufficient!)
condition: the scattering angle on a basic scale must be small
enough, $\theta =l_c/r_L(E) < 1$. It implies the propagation in
the regime with $E > E_c$, where $l_d(E)=l_c \ (E/E_c)^2$.
Considering the low-energy case $E \leq 1\times 10^{18}$~eV, when
the adiabatic energy loss dominates, so that $b(E) = E H_0$ and
\begin{equation} %
E_{\rm max} = E_g(E,t_0) = E e^{H_0t_0} = e E,
\end{equation} %
one readily obtains from Eq~(\ref{r_hor})
\begin{equation} %
r^2_{\rm hor}(E)=\frac{c l_c}{6H_0}\left (\frac{E}{E_c} \right )^2
(e^2-1).
\label{r_h} %
\end{equation} %
Using Eq.~(\ref{horizon}) we get
\begin{equation} %
\frac{E_c}{E} \leq \left [ \frac{c H_0^{-1}}{6 l_c}(e^2-1) \right
]^{-1/2}.
\end{equation} %
From $r_L(E_c) = l_c$ we obtain $E_c = q B_c l_c$, where $q$ is an
electric charge, which equals $300$ for a proton, if $B$ is
measured in Gauss and $E$ in eV.

Finally we have
\begin{equation} %
B_c \leq \frac{E}{q l_c}\left [\frac{c H_0^{-1}}{6 l_c}(e^2-1)
\right ]^{-1/2},
\end{equation} %
or, numerically,
\begin{equation} %
B_c \leq 1.6 \times 10^{-3} \ \frac{E}{10^{17}~{\rm eV}} \left
(\frac{l_c} {1~{\rm Mpc}} \right )^{-1/2} ~{\rm nG}.
\label{B_c} %
\end{equation} %
Therefore, $B_c \leq 1.6 \times 10^{-3}$~nG provides the
quasi-rectilinear propagation for all protons with energies $E
\geq 1\times 10^{17}$~eV, while at lower energies the diffusion
description is applicable. For the reasonably low field $B_c
\approx 0.01$~nG the diffusion becomes valid at  $E  \lesssim
1\times 10^{18}$~eV.

%%%%%%%%%%%%%%%%%%%%%%%%%%%%%%%%%%%%%%%%%%%%%%%%%%%%%%%%%%%%%%%%%%%%%%%%%%%
\section{Diffusive energy spectra of UHECR in the expanding universe}
\label{sec:expandingU} %
In the following calculations we shall use the simplified
illustrative description of magnetic field evolution with
redshift, namely we parametrize the evolution of magnetic
configuration $(l_c,B_c)$ as
\[
l_c(z) = l_c/(1+z), \ \ \   B_c(z)= B_c \ (1+z)^{2-m},
\]
where factor $(1+z)^2$ describes the diminishing of the magnetic
field with time due to magnetic flux conservation and
$(1+z)^{-m}$~ -- due to MHD amplification of the field. The
critical energy $E_c(z)$ found from $r_L(E) = l_c(z)$ is given by
\[
E_c(z)=0.93 \times 10^{18}\ (1+z)^{1-m}\ \frac{B_c}{1~\mbox{nG}}
\]
for $l_c = 1$~Mpc. The maximum redshift used in the calculations
is $z_{\rm max}=4$.

The diffuse flux is calculated for the lattice distribution of the
sources (in the coordinate space $\vec{x}$) with lattice parameter
(the source separation) $d$ and a power-law generation function
for a single source
\begin{equation} %
Q_s(E) = \frac{q_0 (\gamma_g - 2)}{E_0^2} \left( \frac{E}{E_0}
\right)^{-\gamma_g},
\end{equation} %
where $E_0$ is the normalizing energy, for which we will use $1
\times 10^{18}$ eV and $q_0$ has a physical meaning of a source
luminosity in protons with energies $E \geq E_0$, $L_p(\geq E_0)$.
The corresponding emissivity $\mathcal{L}_0 = q_0/d^3$, i.e.\ the
energy production rate in particles with $E \geq E_0$ per unit
comoving volume, will be used to fit the observed spectrum by the
calculated one.

Using to Eqs.~(\ref{solution}--\ref{x_ijk}), one obtains the
diffuse spectrum as
%\begin{equation} %
\begin{eqnarray} %
J_p(E) & = & \frac{c}{4 \pi H_0} \ \frac{q_0 (\gamma_g-2)}{E_0^2}
\times \nonumber \\ %
 & & \sum_i \int \limits_0^{z_g} dz \frac{\left[E_g(E,z)/E_0
\right]^{-\gamma_g}}
  {(1+z)\sqrt{\Omega_m(1+z)^3+ \Lambda}} \
\frac{\exp[-\frac{x_i^2}{4\lambda(E,z)}]}
{[4\pi\lambda(E,z)]^{3/2}} \ \frac{dE_g(E,z)}{dE},
\label{flux-expans} %
\end{eqnarray} %
%\end{equation} %
where summation goes over the sources like in Eqs. (\ref{J_p}) and
(\ref{x_ijk}), the upper limit $z_g$ is provided by maximum energy
of acceleration as $E_g(E,z_g)=E_{\rm max}^{\rm acc}$ or by
$z_{\rm max}$, whichever is smaller, $\gamma_g$ is the generation
index and a formula for $dE_g/dE$ can be found in \cite{BG88} and
\cite{BGG02a}. The analogue of the Syrovatsky variable,
$\lambda(E,z)$, is given by
\begin{equation} %
\lambda(E,z)=\frac{1}{H_0}\int_0^z dz
\frac{(1+z)}{\sqrt{\Omega_m(1+z)^3+ \Lambda}} \ D[E_g(E,z),z].
\label{lambda-expans} %
\end{equation} %

\begin{figure}[ht] %
\centering %
\includegraphics[width=8.0cm]{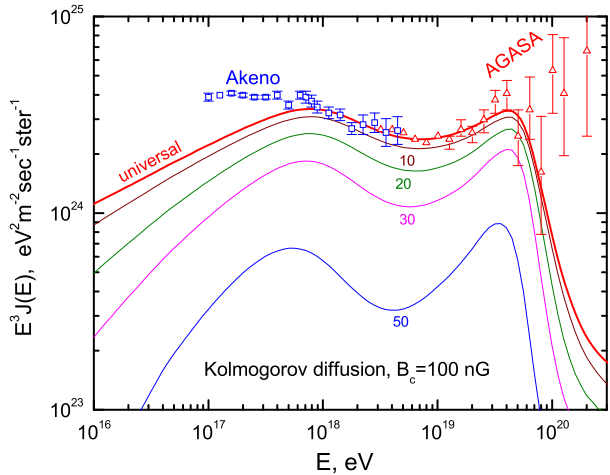} \vspace{2mm}
\caption{Convergence of the diffusive solution (\ref{J_p}) to the
universal spectrum when the distance between sources diminishes
from 50 to 10~Mpc shown by numbers on the curves. The magnetic
field configuration $(B_c,l_c) = (100$~nG, 1~Mpc) with the
Kolmogorov spectrum of turbulence. The emissivity $\mathcal{L}_0 =
q_0/d^3 = 2.4 \times 10^{45}$~erg/Mpc$^3$yr is the same for all
curves. The generation spectrum is $\propto E^{-2.7}$. The
spectrum of Akeno-AGASA \citep{Akeno,AGASA06} is shown for
comparison. }
\label{fig:converg} %
\end{figure} %

First of all we test the correctness of the obtained solution with
the help of propagation theorem \citep{AB04}, which states that
diffusive solution (\ref{J_p}) converges to the universal
spectrum, i.e.\ one calculated for homogeneous source distribution
(see \cite{BGG02a}) when the distances between sources $d
\rightarrow 0$. Fig.~\ref{fig:converg} demonstrates this
convergence for the case of strong magnetic field, $B_c=100$~nG,
and the Kolmogorov diffusion at $E < E_c$. The diffuse fluxes are
calculated using Eq.~(\ref{J_p}) for $d$ diminishing from $50$~Mpc
to $10$~Mpc, keeping the same comoving volume emissivity
$\mathcal{L}_0 = q_0/d^3 = 2.4 \times 10^{45}$~erg/Mpc$^3$yr. The
generation spectrum is given by $\gamma_g=2.7$.  One can observe
the convergence of the calculated spectra to the universal one
when $d \rightarrow 0$, in accordance with propagation theorem.

For small distances between a source and observer, the diffusion
approximation for a particle propagation is not valid. One can see
it from a simple argument that diffusive propagation time $r^2/D$
must be larger than time of rectilinear propagation, $r/c$. This
condition, using $D \sim c \, l_d$, results in $r \lesssim l_d$.
At distances $r \lesssim l_d$ the rectilinear and diffusive
trajectories in magnetic fields differ but little and rectilinear
propagation is a good approximation as far as spectra are
concerned. The number densities of particles  $Q/4\pi c r^2$ and
$Q/4\pi D r$, calculated in rectilinear and diffusive
approximations, respectively, are equal at $r \sim l_d$, where $Q$
is the rate of particle production by a source. At smaller
distances the rectilinear flux dominates, at larger -- the
diffusive one. We calculated the number densities of protons
$n(E,r)$ numerically for both modes of propagations with energy
losses of protons taken into account, and the distance of
transition is taken from equality of the calculated densities. We
know that this recipe is somewhat rough and the interpolation
between two regimes is required, as in calculations by
\cite{AB05}. However any interpolation meets the difficulties by
the following reason.

In fact, the diffusive regime sets up at distances not less than
six diffusion lengths $l_d$. At distances $l_d \lesssim r \lesssim
6\, l_d$ some intermediate regime of propagation in magnetic
fields is valid. When studied in numerical simulations (e.g.\
\cite{Sato03}), the calculated number density $n(E,r)$ satisfies
the particle number conservation $4\pi r^2 n u = Q$, where $u$ is
the streaming velocity, while in case of interpolation there is
only one (unknown) interpolation $n(E,r)$ which conserves the
number of particles. This problem will be studied in detail in
\citep{ABG07}, while for purposes of this paper we can accept the
rough recipe of transition from diffusive to rectilinear regime as
described above. The appearance of the artificial peculiarity
connected with the accepted propagation transition will be useful
serving as a mark for the position of transition in the spectrum.

For rectilinear propagation for the lattice distribution of the
sources the diffuse spectrum is calculated as \citep{BGG02a}
\begin{eqnarray} %
J_p(E) & = & \frac{(\gamma_g-2){\cal L}_0 d}{(4\pi)^2 E_0^2 } \times  \nonumber \\
 &  & \sum \limits_{i,j,k}
\frac{ \left[  E_g(E,z_{ijk})/ E_0
\right]^{-\gamma_g}}{[(i+1/2)^2+(j+1/2)^2+(k+1/2)^2](1+z_{ijk})} \
\frac{dE_g}{dE}(E,z_{ijk}),
\label{flux-lattice}%
\end{eqnarray} %
where $ \mathcal{L}_0 = q_0/d^3 = 2.4 \times 10^{45}
$~erg/(Mpc$^3$yr) is the emissivity, $z_{ijk}$ is the red-shift
for a source with coordinates $i, j, k$, and factor $(1+z_{ijk})$
takes into account the time dilation.
\begin{figure}[ht]
   \begin{minipage}[ht]{8cm}
     \centering
     \includegraphics[width=7.5cm]{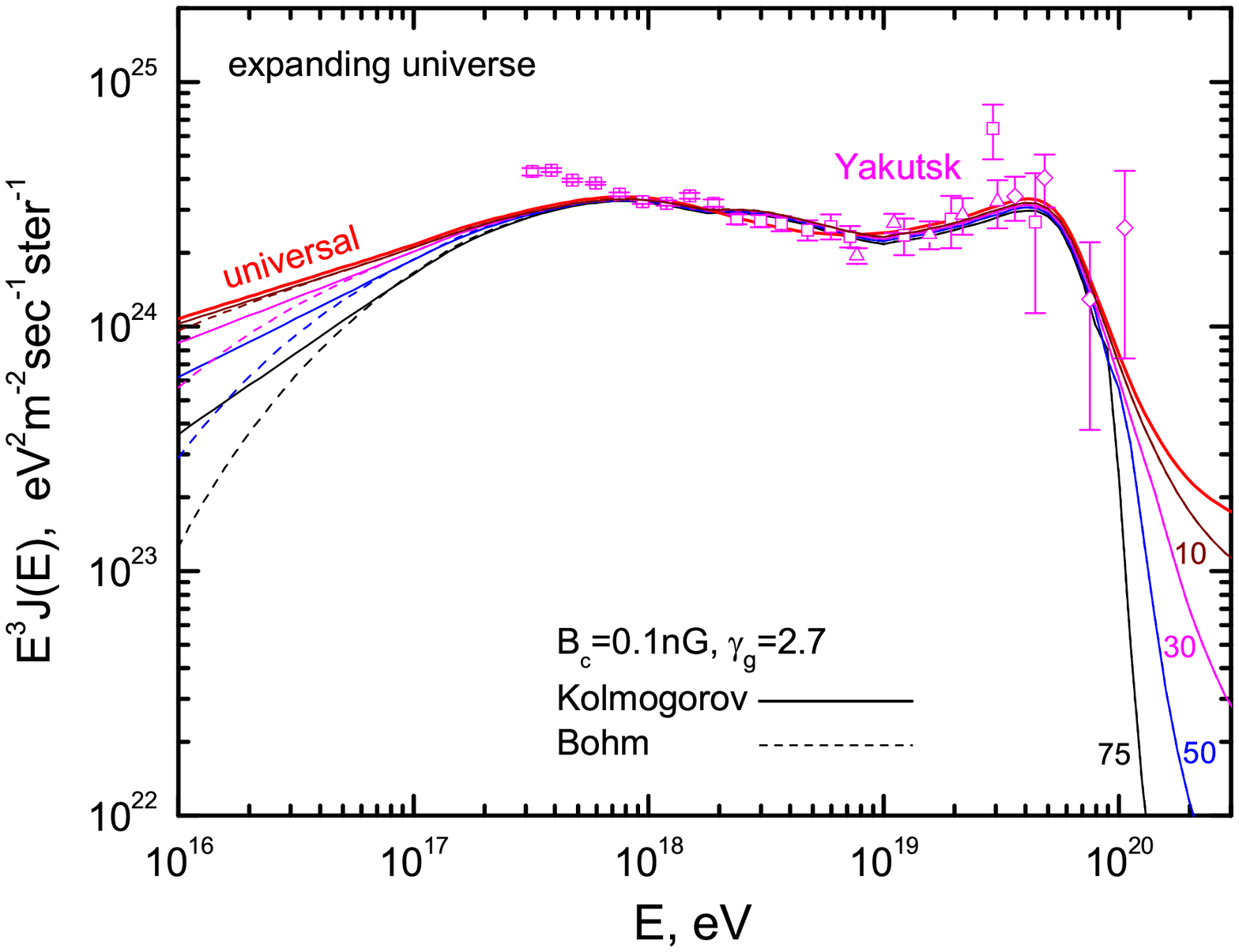}
   \end{minipage}
   \hspace{1mm}
   \vspace{-1mm}
 \begin{minipage}[h]{8cm}
    \centering
    \includegraphics[width=7.5cm]{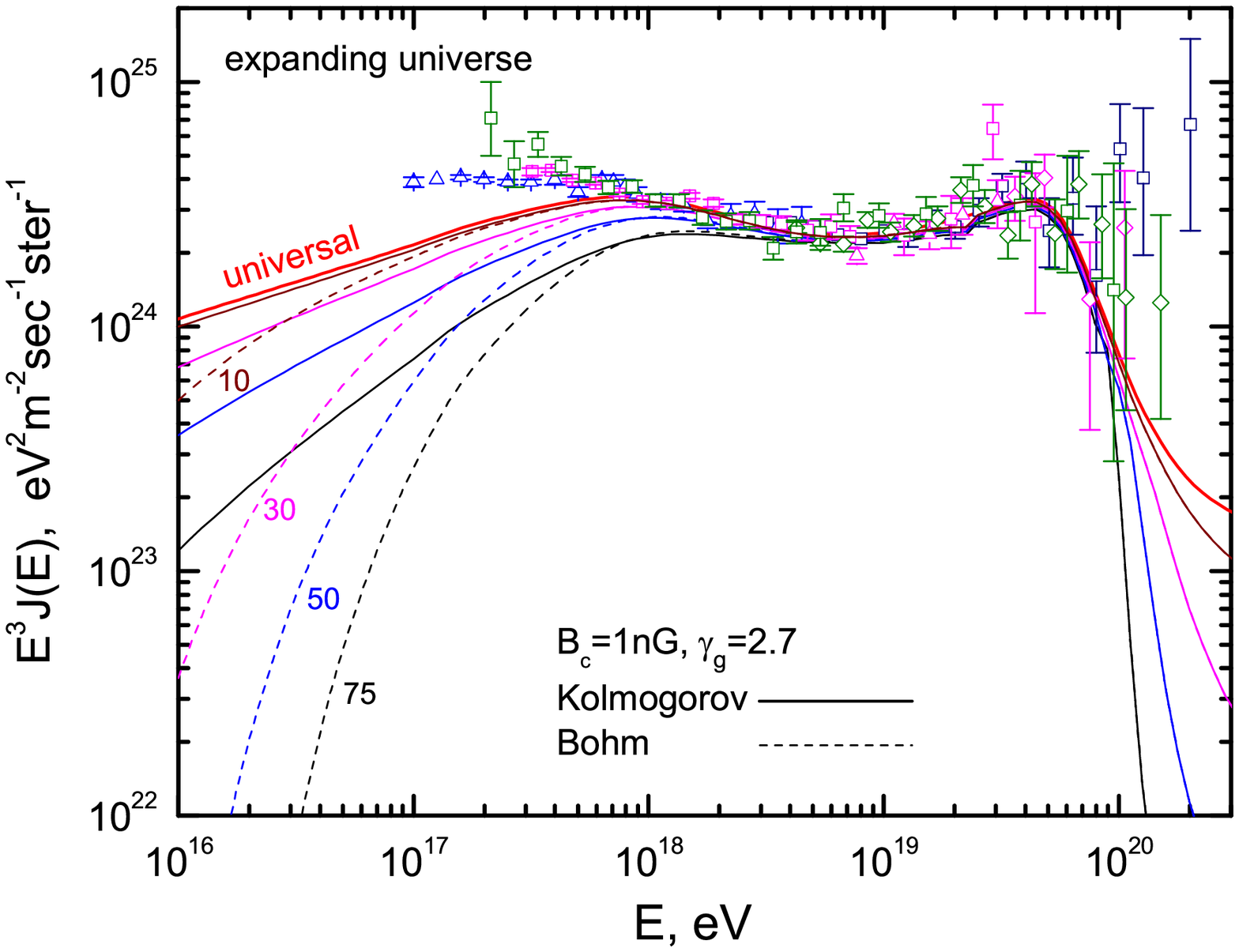}
 \end{minipage}
\caption{Diffusive spectra in the expanding universe for $B_c =
0.1$~nG (left panel) compared with Yakutsk data \citep{Yakutsk}
and $B_c = 1$~nG (right panel) compared with all-data
Akeno-AGASA-HiRes-Yakutsk spectrum
\citep{Akeno,AGASA06,HiRes,Yakutsk}. The spectra are calculated as
a combination of diffusive and rectilinear spectra with the
Kolmogorov (solid lines) and Bohm (dash lines) regimes of
diffusion at low energies $E < E_c$. The numbers at the curves
indicate the separations of the sources in Mpc. The spectral
generation index $\gamma_g=2.7$ and maximum acceleration energy
$E_{\rm max}=1\times 10^{22}$~eV.}
\label{fig:1-0.1-exp} %
\end{figure} %

The calculated spectra in the expanding universe for $B_c=1$~nG
and $B_c=0.1$~nG, both for $m=1$ and $E_{\rm max}=1\times
10^{22}$~eV are shown in Fig.~\ref{fig:1-0.1-exp} in comparison
with all data (left panel) and Yakutsk data \citep{Yakutsk} (right
panel). The all-data spectrum is obtained using the of energy
calibration of all detectors with help of the dip \citep{BGG02a}.
One can observe the peculiarity in predicted spectrum in the right
panel ($B_c=1$~nG) at energy $E \approx 2\times 10^{19}$~eV. This
is the energy of transition to rectilinear propagation. This
peculiarity is unphysical and is connected with the simplified
description of the transition as described above. When magnetic
field diminishes a peculiarity shifts toward lower energy (left
panel) as it should do. At small $d$ the calculated spectra
converge to the universal spectrum, as it must be.
\begin{figure}[ht] %
\centering %
\includegraphics[width=8.0cm]{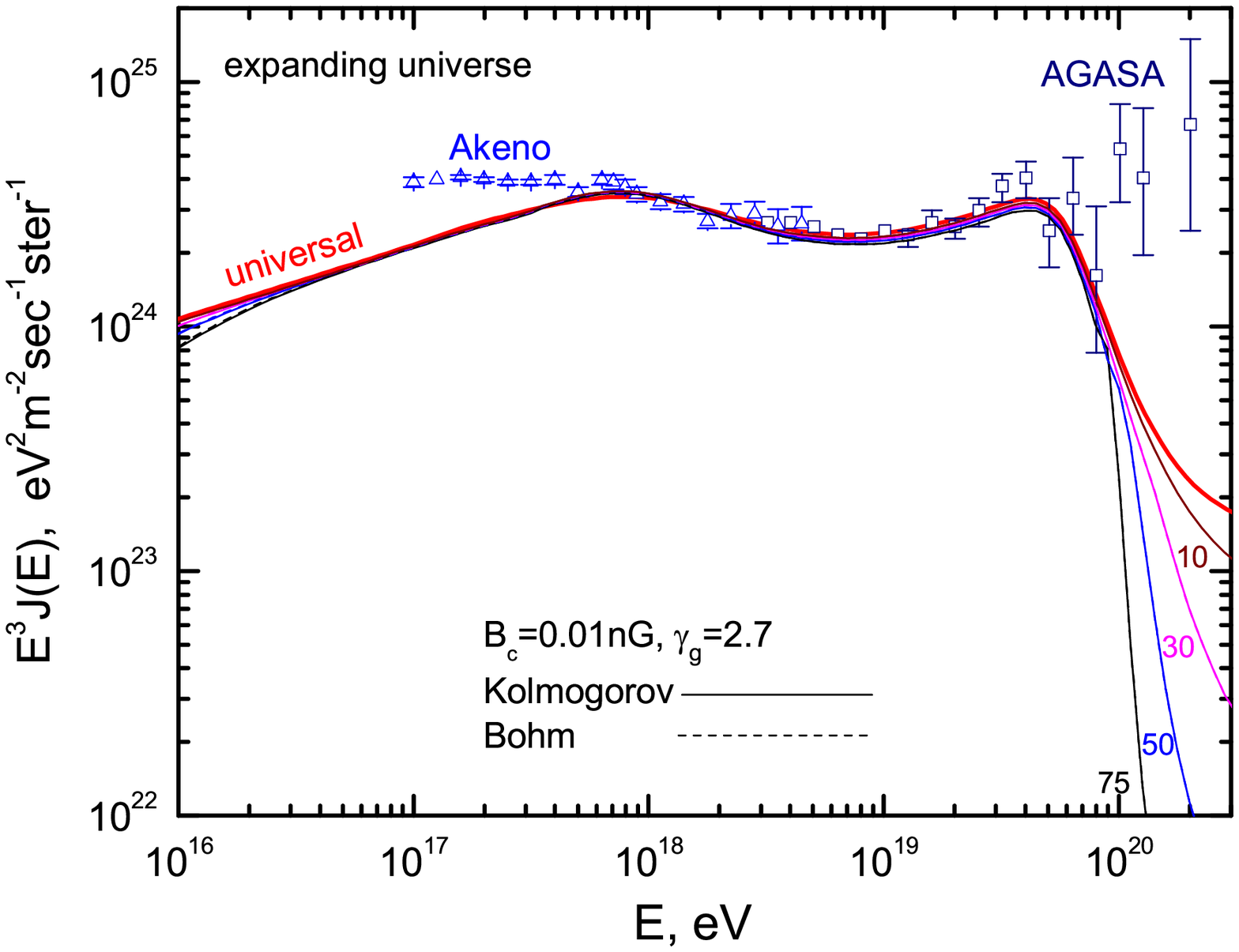}
\vspace{2mm} \caption{The same as in Fig.~\ref{fig:1-0.1-exp}, but
for $B_c=0.01$~nG and Akeno-AGASA \citep{Akeno,AGASA06} data.}
\label{fig:0.01nG-exp} %
\end{figure} %
The spectra for very low magnetic field with $B_c=0.01$~nG is
shown in Fig.~\ref{fig:0.01nG-exp}. One can notice that the
peculiarity is shifted here to the energy $E \sim 2\times
10^{17}$~eV and above this energy the spectrum is rectilinear.
This behavior corresponds to the discussion in
Section~\ref{sec:whydiffusion}.

%%%%%%%%%%%%%%%%%%%%%%%%%%%%%%%%%%%%%%%%%%%%%%%%%%%%%%%%%%%%%%%%%%%%%%%%
\section{Spectra in the static universe}
\label{sec:staticU} %
In this section we study the Syrovatsky solution of UHECR
diffusion equation with aim of comparing it with the BG solution
for the expanding universe. The Syrovatsky solution is valid in
case of infinite space with time-independent diffusion coefficient
$D(E)$ and energy losses $b(E)$. This implies the static universe.

We define the static universe in which the stationary diffusion
equation with the Syrovatsky solution is embedded in the following
way. There is no expansion. The Hubble constant is assumed as a
formal parameter $H_0 = 72$ km/s Mpc, which defines the "age"
$t_0$ of the universe according to WMAP relation $H_0 t_0 =
0.993$, so that $c t_0$ is the size of the universe: the space
density of the UHECR sources outside the sphere of radius $c t_0$
is $n_s = 0$. The temperature of CMB photons is constant, $T_0 =
2.728$~K, and thus the interaction energy losses $dE/dt = -b(E)$
are time-independent.

Apart from interaction energy losses, we assume the fictitious
adiabatic energy losses described by $dE/dt = -H_0 E$. In this
approach we follow the work by \cite{AB05}.

The universal spectrum in the static universe is different from
that in the expanding universe. It is defined in the same way as
in the expanding universe \citep{BGG02a}, namely from the number
of particles conservation
\begin{equation} %
n_{\rm univ}(E)=\int^{t_0}_0 dt\, Q_{\rm gen}[E_g(E,t)]\,
\frac{dE_g}{dE},
\label{eq:conserv} %
\end{equation} %
where $Q_{\rm gen}(E_g)$ is the generation rate per unit volume
and $E_g(E,t)$ is determined by the evolution equation $dE/dt =
-b(E)$. Note that in the expanding universe $Q_{\rm gen}(E,t)$ and
$n(E,t)$ are related to a comoving volume. In contrast to the
expanding universe, the ratio $dE_g/dE$ in Eq.~(\ref{eq:conserv})
is given by a simple expression \citep{BG88} $dE_g/dE =
b(E_g)/b(E)$. Using
\begin{equation} %
Q_{\rm gen}(E_g)=\frac{(\gamma_g-2)\mathcal{L}_0}{E_0^2} \left(
\frac{E_g}{E_0} \right)^{-\gamma_g},
\label{eq:Q} %
\end{equation} %
where $\mathcal{L}_0$ is the emissivity in particles with energies
$E \geq E_0$ for spectral index $\gamma_g > 2$, one easily obtains
the universal spectrum in analytic form:
\begin{equation} %
n_{\rm univ}(E) = \frac{\gamma_g-2}{\gamma_g-1} \
\frac{\mathcal{L}_0}{E_0^2} \ \frac{\left(E/E_0 \right)^{-\gamma_g
}}{E^{-1}b(E)} \left [1- \left (\frac{E}{E_{\rm max}}\right
)^{\gamma_g-1} \right ],
\label{eq:univ} %
\end{equation} %
where $E_{\rm max}={\rm min}[E_g(E,t_0), E_{\rm max}^{\rm acc}]$.
For the diffuse spectra calculations we use the lattice
distribution of the sources like in the previous section, with the
same procedure of transition from diffusive to rectilinear
propagation, but using for diffusive propagation the Syrovatsky
solution (see also \cite{AB05}), instead of the BG solution.

\begin{figure}[ht] %
   \begin{minipage}[ht]{8cm}
     \centering
     \includegraphics[width=7.5cm]{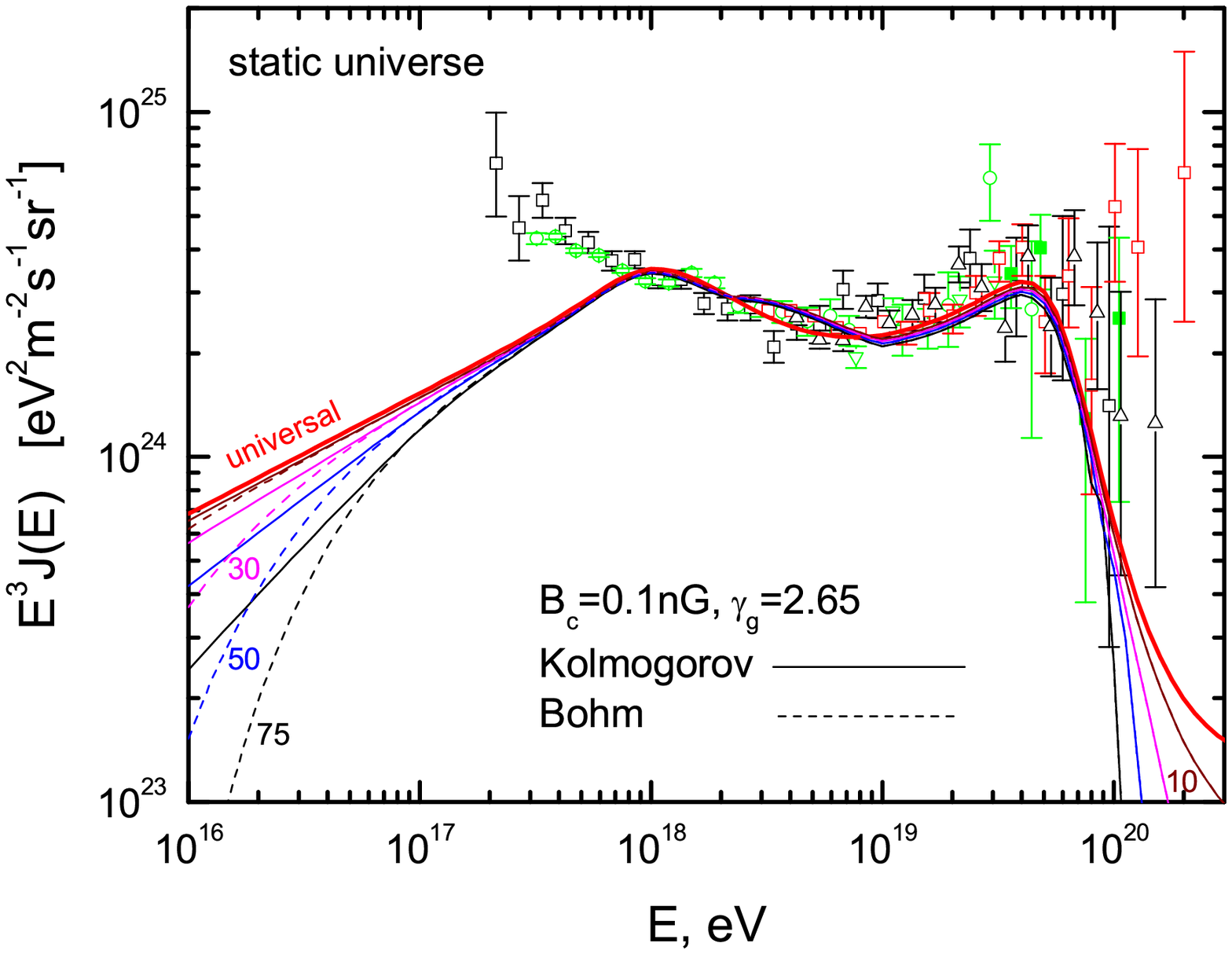}
   \end{minipage}
   \hspace{1mm}
   \vspace{-1mm}
 \begin{minipage}[h]{8cm}
    \centering
    \includegraphics[width=7.5cm]{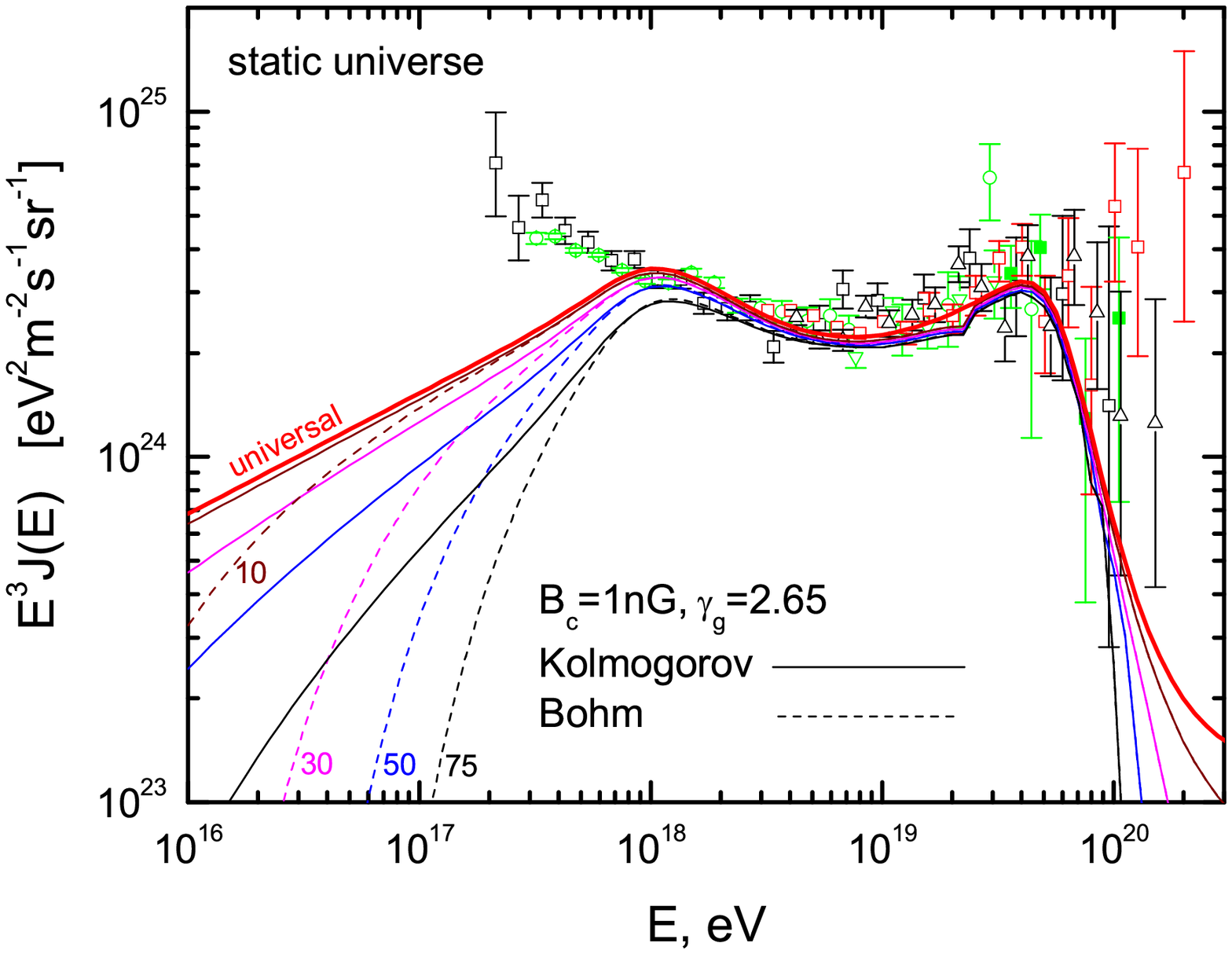}
 \end{minipage}
\caption{Diffuse spectra in the static universe for $B_c = 0.1$~nG
  (left panel) and $B_c = 1$~nG (right panel). The spectra are
  calculated as combination of diffusive and rectilinear spectra
  with the Kolmogorov (solid lines) and Bohm (dash lines) regimes
  of diffusion at low energies $E < E_c$. The numbers at the
  curves indicate the separation of the sources in Mpc. Observational
  data are the same as in the right panel of Fig.~\ref{fig:1-0.1-exp}.}
\label{fig:1-0.1-stat} %
\end{figure}

In Fig.~\ref{fig:1-0.1-stat} we present the diffuse UHECR spectra
calculated, like in Section~\ref{sec:expandingU}, in the grid
model with spacing $d$. UHECR sources are located in vertices of
the grid. The spectra are calculated as combination of the
diffusive flux, described by the Syrovatsky solution, combined
with the rectilinear flux. They are presented for two magnetic
field configurations $(B_c, l_c)$ equal to $(1\mbox{~nG},
1\mbox{~Mpc})$ and $(0.1\mbox{~nG}, 1\mbox{~Mpc})$ and for
different spacings $d$ indicated by the numbers in Mpc at the
corresponding curves. Like in Section~\ref{sec:expandingU}, the
spectra have irregularities at energy of transition from diffusive
to rectilinear propagation, caused by the rough method of sewing
together of the two propagation regimes.

The calculated diffuse fluxes converge to the universal spectrum
(\ref{eq:univ}) when $d \rightarrow 0$, as it should be according
to propagation theorem. In contrast to calculations by
\cite{AB05}, we have obtained the best fit of the data using
$\gamma_g=2.65$, which is insignificantly different from $\gamma_g
= 2.7$ of \citep{AB05}. One can see the basic agreement of these
spectra with those in the expanding universe, including spectrum
peculiarities caused by transition from diffusive to rectilinear
propagation.

%%%%%%%%%%%%%%%%%%%%%%%%%%%%%%%%%%%%%%%%%%%%%%%%%%%%%%%%%%%%%%%%%%%%%%%
\section{Comparison of spectra in expanding and static universes}
\label{sec:comparison}%
The direct comparison of the BG and Syrovatsky solutions of
diffusion equations is not possible because they are embedded in
the different cosmological environments. While the BG solution is
valid for the expanding universe, the Syrovatsky solution is valid
only for the static universe. Using two different cosmological
models for these solutions, there are two ways of comparison.

The first one is given by equal values of parameters in both
solutions. In this method for BG solution we use the standard
cosmological parameters for expanding universe $H_0$, $\Omega_m$,
$\Lambda$ and maximum red-shift $z_{max}$ up to which UHECR
sources are still active, magnetic field configuration ($B_c,
l_c$), separation $d$ and UHECR parameters $\gamma_g$ and
$\mathcal{L}_0$, determined by the best fit of the observed
spectrum. For the static universe with the Syrovatsky solution we
use the same parameters $H_0$, $d$, ($B_c, l_c$), $\gamma_g$ and
$\mathcal{L}_0$. The maximum red-shift in the BG solution we chose
as that providing the age of universe which equals to $t_0 =
H_0^{-1}$ in the static universe ($z_{max}=1.465$). This formal
method of comparison will be referred to as "equal-parameter
method".

Physically better justified comparison is given by the best fit
method, in which $\gamma_g$ and $\mathcal{L}_0$ are chosen as the
best fit parameters for the both solutions, respectively. As a
matter of fact, we would use the best-fit parameters $\gamma_g$
and $\mathcal{L}_0$ for each solution, if we considered them
independently. The weakness of this method is a "forced" agreement
at the energy range of measured UHECR spectrum, provided by the
best-fit parameters to the same spectrum. Then the difference of
the two solutions becomes most appreciable at $E \leq 1\times
10^{18}$ eV.

\begin{figure}[ht]
   \begin{minipage}[ht]{8cm}
     \centering
     \includegraphics[width=7.5cm]{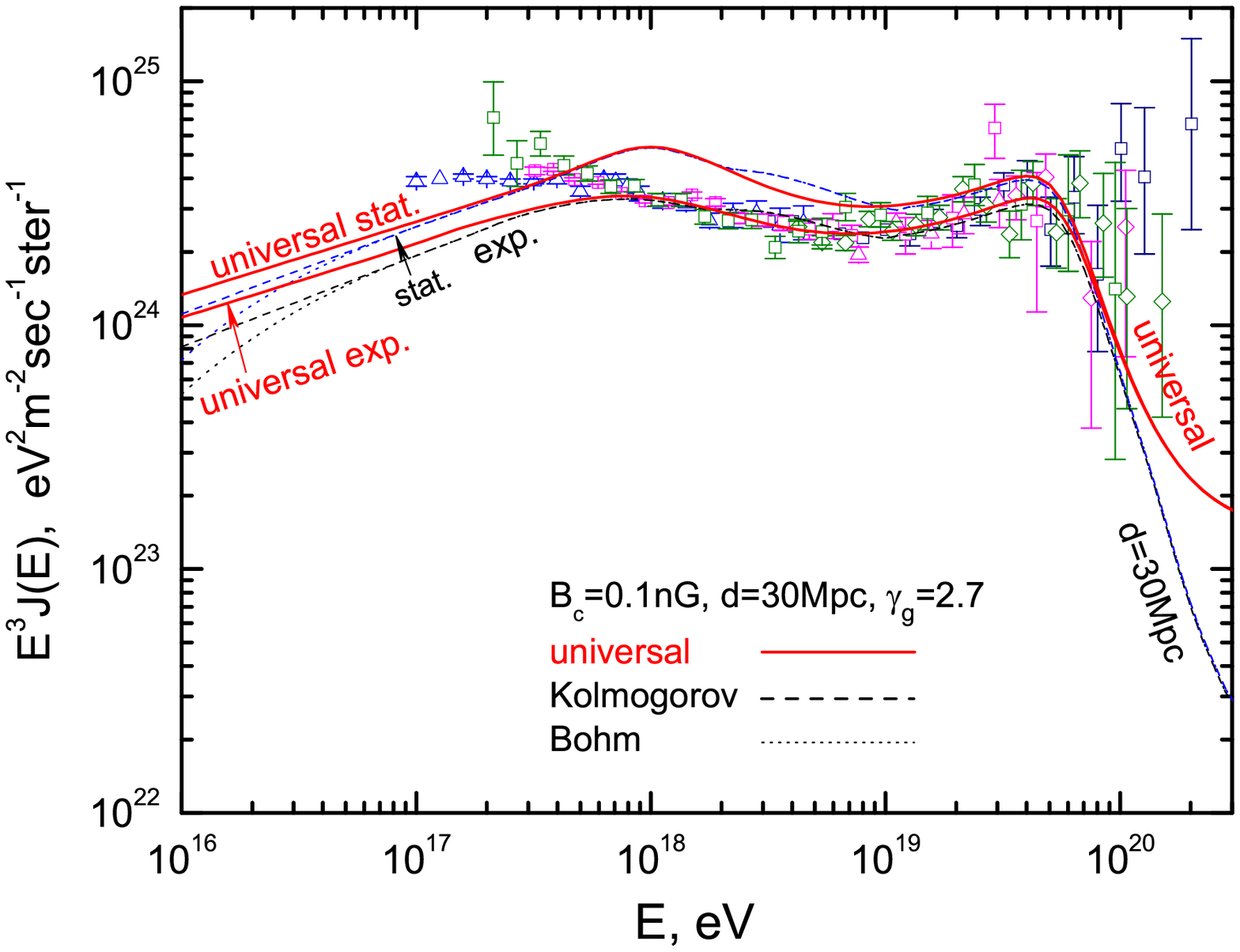}
   \end{minipage}
   \hspace{1mm}
 \begin{minipage}[h]{8cm}
    \centering
    \includegraphics[width=7.5cm]{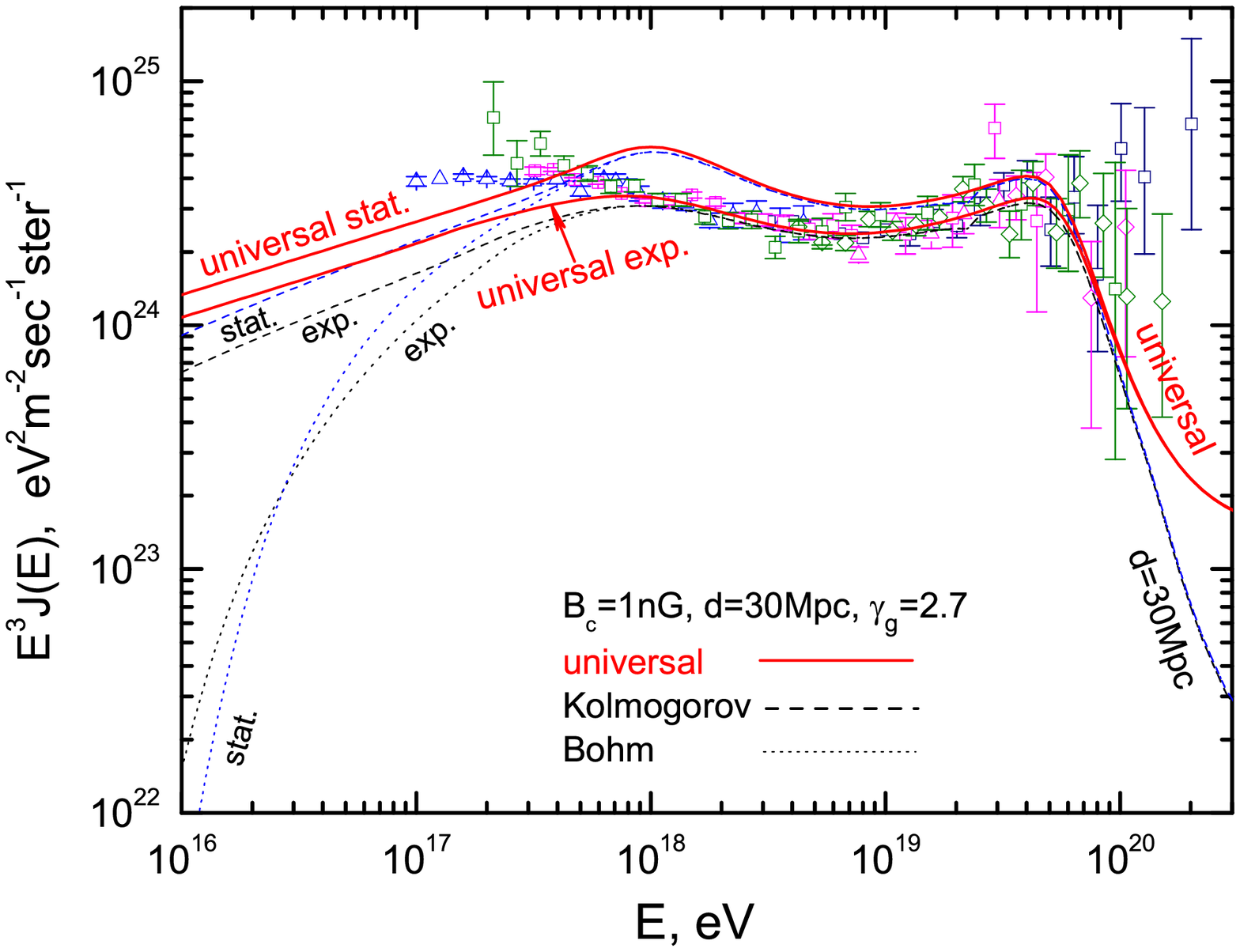}
 \end{minipage}%
\caption{Equal-parameter comparison of the expanding (exp.) and
  static (stat.) universe solutions for $\gamma_g=2.7$,
  $\mathcal{L}_0 = 2.4 \times 10^{45}$~erg/Mpc$^3$yr and $d =
  30$~Mpc. The left panel shows the case $B_c = 0.1$~nG, the right
  panel is for $B_c = 1$~nG. The universal spectra are shown by
  solid lines. The cases of the Kolmogorov and Bohm diffusion at
  $E < E_c$ are shown by dash and dot lines, respectively. Data in
  both panels are the same as in the right panel of
  Fig.~\ref{fig:1-0.1-exp}.
        } %
\label{fig:ESB1g27} %
\end{figure} %
Equal-parameter comparison of the expanding and static universe
solutions are shown in Fig.~\ref{fig:ESB1g27} for $\gamma_g =
2.7$, $\mathcal{L}_0 = 2.4 \times 10^{45}$~erg/Mpc$^3$yr and
$d=30$~Mpc. In the left panel $B_c = 0.1$~nG  and in the right
panel $B_c = 1$~nG. The universal spectra for the expanding and
static universes are shown by solid lines. The Kolmogorov
diffusion at $E<E_c$ is presented by dash lines and the Bohm
diffusion -- by dot lines. Note that universal spectra are
different at $E < 1\times 10^{19}$~eV as they must be due to
excessive energy losses in the expanding universe caused by the
increasing of CMB temperature $T(z)$ with red-shift. At the
highest energy end $E \gtrsim 6 \times 10^{19}$ eV the two
solutions coincide exactly because at these energies the
energy-attenuation time is short, the CMB temperature does not
change during time-of-flight and the expanding universe case
becomes static. As one can see in Fig.~\ref{fig:ESB1g27}, at these
energies both universal spectra are the same and all spectra for
$d=30$ Mpc merge into one curve. The difference of fluxes at lower
energies is naturally explained by the increase of energy losses
in the expanding universe because of $T_{CMB}(z)$ dependence, and
due to cosmological evolution of the magnetic field.

\begin{figure}[ht]
  \begin{minipage}[ht]{8cm}
     \centering
     \includegraphics[width=7.5cm]{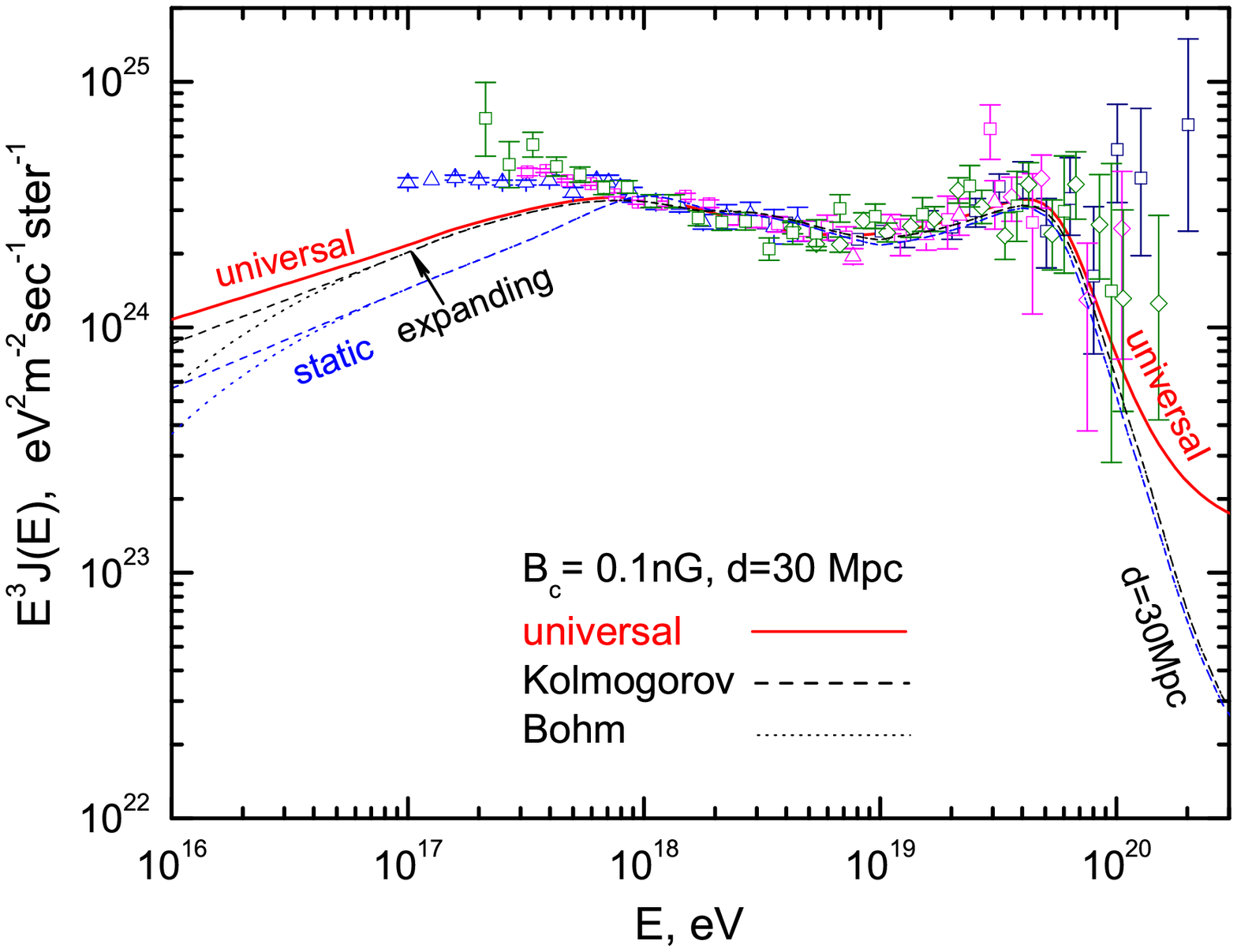}
   \end{minipage}
   \hspace{1mm}
 \begin{minipage}[h]{8cm}
    \centering
    \includegraphics[width=7.5cm]{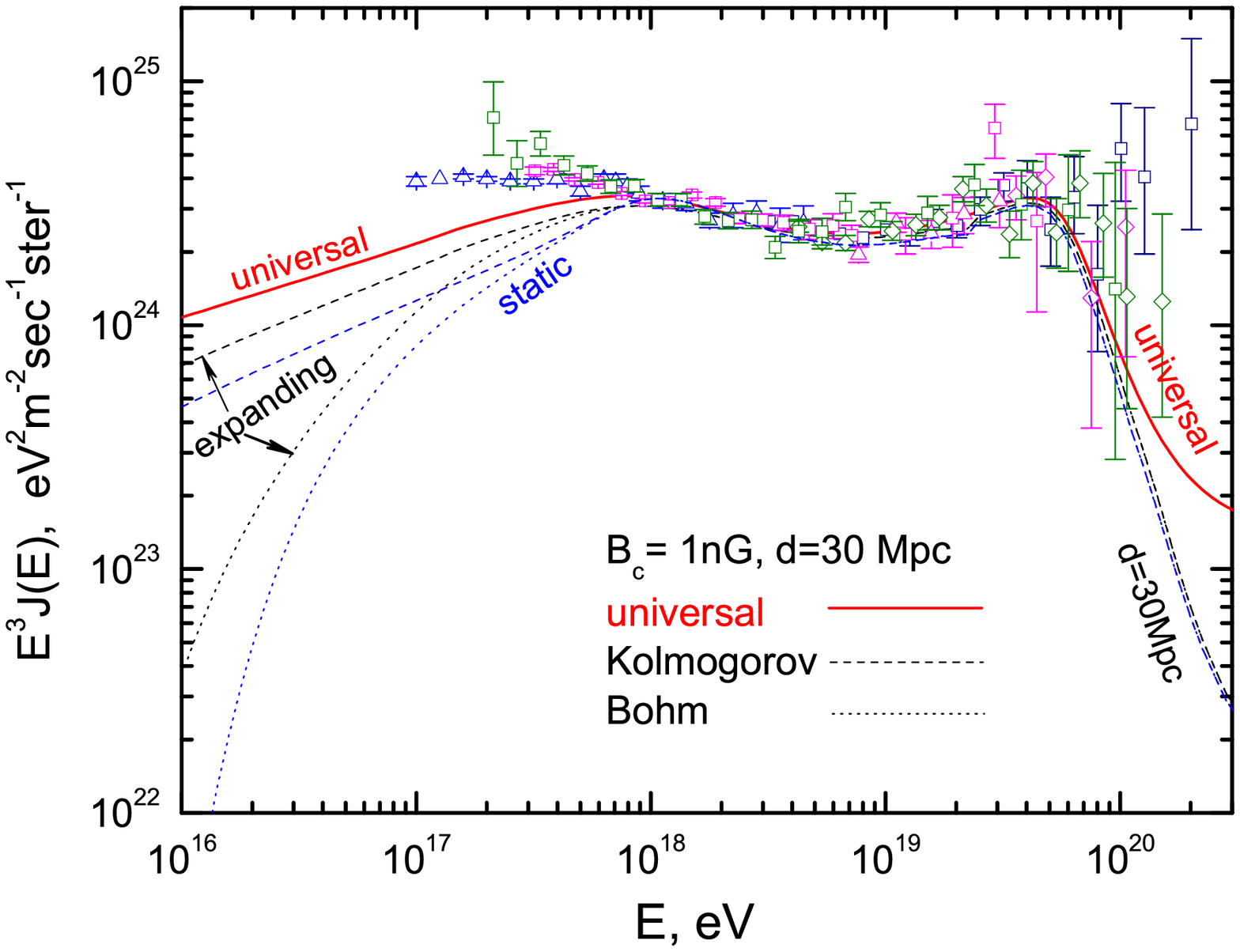}
 \end{minipage} %
\caption{Best fit comparison of the expanding and static universe
  solutions for $d=30$ Mpc. Best-fit parameters are $\gamma_g=2.7,
  \mathcal{L}_0 = 2.4 \times 10^{45} \ \mbox{erg}/\mbox{Mpc}^3
  \mbox{yr})$ for expanding and $\gamma_g=2.65, \mathcal{L}_0 =
  5.7 \times 10^{44} \ \mbox{erg}/\mbox{Mpc}^3 \mbox{yr})$ for
  static solutions, respectively. The universal spectrum is shown
  for the expanding universe. The left panel exposes the case of
  $B_c = 0.1$ nG and the right one the case of $B_c = 1$ nG. The
  combined data of all arrays are shown for comparison. Data in
  both panels are the same as in the right panel of
  Fig.~\ref{fig:1-0.1-exp}. } %
\label{fig:ESB1} %
\end{figure} %
The best-fit comparison of the expanding and static universe
solutions is shown in Fig.~\ref{fig:ESB1}. The best-fit parameters
are $(\gamma_g=2.7, \mathcal{L}_0 = 2.4 \times 10^{45} \
\mbox{erg}/\mbox{Mpc}^3 \mbox{yr})$ and $(\gamma_g=2.65,
\mathcal{L}_0 = 5.7 \times 10^{44} \ \mbox{erg}/\mbox{Mpc}^3
\mbox{yr})$ for the expanding and static universes, respectively.
The left panel shows the case $B_c = 0.1$~nG and the right panel
$B_c = 1$~nG, both for $d=30$~Mpc. One can see a good (though
"forced") agreement between the expanding universe and static
universe solutions in the energy range of UHECR observations,
which is improved in comparison with equal-parameter method
because of the choice of the best-fit parameters
$(\gamma_g,\mathcal{L}_0)$ different for each case.

However, we emphasize again that this is a natural way of
selection of the solution in independent analysis. At energies
below $1\times 10^{18}$~eV the larger discrepancies are seen,
being induced by larger energy losses and evolution of magnetic
field in case of expanding universe.

In conclusion, one can see a reasonably good agreement between the
Syrovatsky solution, embedded in static universe model, with the
BG solution for expanding universe at energies $E > 1 \times
10^{18}$~eV with noticeable discrepancies at smaller energies,
which are natural and understandable.
%%%%%%%%%%%%%%%%%%%%%%%%%%%%%%%%%%%%%%%%%%%%%%%%%%%%%%%%%%%%%%%%%%%%%%%
\section{Conclusions}
\label{sec:conclusions} %
In this paper we study the application of solution of diffusion
equation in expanding universe obtained in the paper~I to the
propagation of UHE protons. However, we do not consider here the
detailed picture with realistic evolution of magnetic field in
expanding universe, and with a realistic transition from the
diffusive to the quasi-rectilinear propagation. It is to be
considered in our next work \cite{ABG07}.

In this paper we demonstrate that solution of diffusion equation
found in paper~I looks quite reasonable when applied to realistic
models. Numerically these solutions are similar to the Syrovatsky
solutions valid for the static universe with the understandable
distinctions at low energies.

The diffusion spectra in expanding universe are presented in
Figs.\ \ref{fig:converg}, \ref{fig:1-0.1-exp},
\ref{fig:0.01nG-exp} for the following magnetic configurations
$(B_c,l_c) = (100 \mbox{~nG}, 1 \mbox{~Mpc})$, $(1 \mbox{~nG}, 1
\mbox{~Mpc})$, $(0.1 \mbox{~nG}, 1 \mbox{~Mpc})$ and $(0.01
\mbox{~nG}, 1 \mbox{~Mpc})$, respectively. In the latter case the
energy spectra at $E > 1 \times 10^{17}$ eV are practically the
same as in the case of rectilinear propagation. The evolution of
magnetic field in the expanding universe is given by one
illustrative example. The spectra are shown for different
separations $d$ of the sources. Transition from diffusive to
rectilinear propagation is described by the simplified recipe,
which results in the artificial spectral features as described in
Section \ref{sec:expandingU}. The best fit of the observed
spectrum needs the same index of generation spectrum $\gamma_g =
2.7$ as in the case of universal spectrum. The diffusive spectrum
converges to universal spectrum when $d \rightarrow 0$, as it
should be according to the propagation theorem.

The Syrovatsky solution of the diffusion equation, i.e.\ one when
the diffusion coefficient $D(E)$ and energy losses $dE/dt = -
b(E)$ do not depend on time $t$, is valid only for the static
universe. In the static universe we make several additional
assumptions. We introduce the Hubble constant $H_0$ as a formal
parameter, which determines the fictitious "adiabatic" energy
losses $dE/dt = -E H_0$. The "age" of the universe $t_0 =
H_0^{-1}$ determines fact the sphere of radius $r_0 = c t_0$
occupied by the sources. For this universe we calculate the
universal spectrum given by Eq.~(\ref{eq:univ}) and the diffusive
spectra for given magnetic configuration $(B_c, l_c)$ and
different separations of the sources $d$. The convergence to the
universal spectrum occurs when $d \rightarrow 0$ as it should be.
The best fit of the observational data is obtained when $\gamma_g
= 2.65$, which can be considered as a good agreement with the work
by \cite{AB04}, where $\gamma_g = 2.7$ was used.

For the comparison of the BG and Syrovatsky solutions we use two
schemes. In the formal one we compare the BG and Syrovatsky
solutions for the same magnetic configurations $(B_c, l_c)$ and
the same emissivities $\mathcal{L}_0$. One can see from
Fig.~\ref{fig:ESB1g27} that both solutions coincide exactly at $E
\geq 6 \times 10^{19}$ eV, when effect of universe expansion (most
notably variation of CMB temperature) can be neglected. At lower
energies the difference in the spectra naturally emerges due to
$T_{CMB}(t)$ dependence and thus due to different energy losses in
the two solutions. Since the energy losses in the expanding
universe are larger, the BG spectra occur below the Syrovatsky
spectra.

For practical applications the discrepancy in the spectra at
energy range $1 \times 10^{18}-5 \times 10^{19}$ eV is not
essential, because it is eliminated by renormalization of the
calculated flux, i.e.\ by changing the emissivity $\mathcal{L}_0$
for the static universe solution. In fact, this procedure is
necessary for fitting of the observed spectra.

The comparison of two solutions as given above is formal. As was
emphasized above, the Syrovatsky solution, which is valid for
infinite space and time, with time-independent $D(E)$ and $b(E)$,
needs the specific definition of the static universe. Only in this
case one obtains the physically viable solution. This solution
needs the best fit parameters $\gamma_g$ and $\mathcal{L}_0$,
which are different from those in expanding universe. It is
physically more meaningful to compare the spectra using for static
universe its own best fit parameters  $\gamma_g$ and
$\mathcal{L}_0$ different from that in the expanding universe. The
comparison shown in Figs.~\ref{fig:ESB1} reveals less
discrepancies than in the formal scheme of comparison.

%%%%%%%%%%%%%%%%%%%%%%%%%%%%%%%%%%%%%%%%%%%%%%%%%%%%%%%%%%%%%%%%%%%%%%%
\section{Acknowledgments}
We are grateful to Vitaly Kudryavtsev for valuable discussions and
participation in a part of this work. This work has been supported
by TA-DUSL activity of the ILIAS program (contract No.\
RII3-CT-2004-506222) as part of the EU FP6 programme.

%%%%%%%%%%%%%%%%%%%%%%%%%%%%%%%%%%%%%%%%%%%%%%%%%%%%%%%%%%%%%%%%%%%%%%%


\begin{thebibliography}{dummy}

\bibitem[Abbasi \etal(2004)]{HiRes}
Abbasi, R.U. \etal, [HiRes Collaboration] 2004, \prl, 92, 151101

\bibitem[Aloisio \&\ Berezinsky(2004)]{AB04} Aloisio, R., \& Berezinsky, V.
2004, \apj, 612, 900

\bibitem[Aloisio \&\ Berezinsky(2005)]{AB05} Aloisio, R., \& Berezinsky, V.
2005, \apj, 625, 249

\bibitem[Aloisio et al.(2007a)]{Aloisioetal} Aloisio, R., Berezinsky, V.,
Blasi, P., Gazizov, A., Grigorieva, S., \& Hnatyk, B. 2007a,
Astropart.\ Phys., 27, 76

\bibitem[Aloisio et al.(2007b)]{ABG07} Aloisio, R., Berezinsky, V., \& Gazizov
A.Z. 2007b, in progress

\bibitem[Berezinsky \& Grigorieva(1988)]{BG88}
Berezinsky, V.S., \& Grigorieva, S.I. 1988, \aap, 199, 1

\bibitem[Berezinsky et al.(1990)]{book}
Berezinsky, V.S., Bulanov, S.V., Dogiel, V.A., Ginzburg, V.L., \&
Ptuskin, V.S. 1990, Astrophysics of Cosmic Rays, North-Holland.

%\bibitem[Berezinsky et al.(1990b)]{Bere90b}
%Berezinsky, V.S., Dogiel, V.A., and Grigorieva, S.I. 1990b, \aap,
%232, 582

\bibitem[Berezinsky et al.(2002a)]{BGG02a}
Berezinsky, V., Gazizov, A.Z., \& Grigorieva, S.I. 2002a, \prd,
74, 043005; [hep-ph/0204357 v1]

\bibitem[Berezinsky et al.(2002b)]{BGG02b}
Berezinsky, V., Gazizov, A.Z., \& Grigorieva, S.I. 2002b,
astro-ph/0210095

\bibitem[Berezinsky \& Gazizov(2006)]{BG06}
Berezinsky, V., \& Gazizov, A.Z. 2006, \apj, 643, 8

\bibitem[Blasi, Burles \& Olinto(1999)]{BBO99}
Blasi, P., Burles, S. \& Olinto, A. 1999 \apj, 514, L79

\bibitem[Egorova \etal(2004)]{Yakutsk}
Egorova, V.P., \etal, [Yakutsk Collaboration] 2004, Nucl.\ Phys.\
B (Proc.\ Suppl.), 136, 3

\bibitem[Honda \etal(1993)]{Akeno}
Honda, M., \etal, [Akeno Collaboration] 1993, \prd, 70, 525

\bibitem[Lemoine(2005)]{Lemoine}
Lemoine, M. 2005, \prd, 71, 083007

\bibitem[Parizot(2004)]{Parizot04}
Parizot, E. 2004, Nucl.\ Phys.\ B (Proc.\ Suppl.) 136, 169

%\bibitem[Kolb \& Turner(1990)]{KT} Kolb, E.W., \& Turner, M.S. 1990,
%The Early Universe, Westview Press

\bibitem[Shinozaki(2006)]{AGASA06}
Shinozaki, K., [AGASA Collaboration] 2006,  Nucl.\ Phys.\ B
(Proc.\ Suppl.), 151, 3

\bibitem[Syrovatskii(1959)]{Syrov}
Syrovatskii, S.I. 1959, Sov.\ Astron.\ J., 3, 22 [1959, Astron.\
Zh., 36, 17]

\bibitem[Yoshiguchi(2003)]{Sato03}
Yoshiguchi, H., Nagataki, S., Tsubaki, S. \& Sato, K. 2003, \apj,
586, 1211

%\bibitem[Lifshitz \& Pitaevskii(2001)]{Lifshi} Lifshitz, E.M., \&
%Pitaevskii, L.P. 2001, Physical Kinetics (Vol.\ 10 of Theoretical
%Physics by Landau, L.D., \& Lifshitz, E.M.) Fizmatlit

%\bibitem[Peebles(1980)]{Peebles}
%Peebles, P.J.E. 1980, The Large-Scale Structure of the Universe,
%Princeton Series in Physics

%\bibitem[Weinberg(1972)]{Weinberg}
%Weinberg, S. 1972, Gravitation and Cosmology, John Wiley and Sons

\end{thebibliography}
\end{document}